\documentclass[11pt]{article}

\usepackage[T1]{fontenc}
\usepackage[utf8]{inputenc}
\usepackage{lmodern}
\usepackage[margin=1in]{geometry}
\usepackage{microtype}
\usepackage{graphicx}
\usepackage{booktabs,longtable,array,calc}
\usepackage{pdflscape}
\usepackage[super,sort&compress]{natbib}
\usepackage[hidelinks]{hyperref}
\usepackage{xurl}
\usepackage{caption}
\usepackage{placeins}

\renewcommand{\arraystretch}{1.12}
\makeatletter
\def\maxwidth{\ifdim\Gin@nat@width>\linewidth\linewidth\else\Gin@nat@width\fi}
\def\maxheight{\ifdim\Gin@nat@height>0.88\textheight 0.88\textheight\else\Gin@nat@height\fi}
\makeatother
\setkeys{Gin}{width=\maxwidth,height=\maxheight,keepaspectratio}

\hypersetup{
  pdftitle={Delusions and Harms Associated With AI Chatbot Use: Early Evidence From 185 Real-World Reports},
  pdfauthor={Hamilton Morrin et al.},
  pdfsubject={AI chatbot use and reported mental health harms}
}

\title{Delusions and Harms Associated With AI Chatbot Use: Early Evidence From 185 Real-World Reports}
\author{\normalsize Hamilton Morrin\textsuperscript{1,2,3,4},
\normalsize Vinitha Soundararajan\textsuperscript{2},
\normalsize Thomas Cheliotis-James\textsuperscript{2,3}\\
\normalsize Boris Warszawski\textsuperscript{2,3},
\normalsize Joshua Fakulujo\textsuperscript{5},
\normalsize Zeqi Jia\textsuperscript{1,3},
\normalsize Etienne Brisson\textsuperscript{6},
\normalsize Thomas A. Pollak\textsuperscript{1,2,3}}
\date{}

\begin{document}
\maketitle
\vspace{-2.2em}

{\small
\begin{list}{}{%
  \setlength{\leftmargin}{1.5em}%
  \setlength{\labelwidth}{1.1em}%
  \setlength{\labelsep}{0.4em}%
  \setlength{\rightmargin}{0pt}%
  \setlength{\itemsep}{0pt}%
  \setlength{\parsep}{0pt}%
  \setlength{\topsep}{0pt}}
\item[\textsuperscript{1}] LAMBDA (Looking After Minds and Brains in the Digital Age), Institute of Psychiatry, Psychology, \& Neuroscience, King's College London, London, UK.
\item[\textsuperscript{2}] South London and Maudsley NHS Foundation Trust, London, UK.
\item[\textsuperscript{3}] Department of Psychosis Studies, Institute of Psychiatry, Psychology, \& Neuroscience, King's College London, London, UK.
\item[\textsuperscript{4}] Department of Psychological Medicine, Institute of Psychiatry, Psychology, \& Neuroscience, King's College London, London, UK.
\item[\textsuperscript{5}] Faculty of Life Sciences and Medicine, King's College London, London, UK.
\item[\textsuperscript{6}] The Human Line Project, Montreal, Quebec, Canada.
\end{list}}

\vspace{0.6em}
\textbf{Corresponding author:} Dr Hamilton Morrin, \href{mailto:hamilton.morrin@kcl.ac.uk}{hamilton.morrin@kcl.ac.uk}

{\setlength{\parskip}{3pt plus 1pt minus 0.5pt}
\begin{abstract}

\textbf{Importance:} Reports have raised concerns that generative
artificial intelligence (AI) chatbots may validate or elaborate
delusional beliefs, respond inappropriately to suicidal ideation, and
contribute to mental health harms, but real-world data on reported harms
remain limited.

\textbf{Objective:} To characterize psychopathological features, chatbot
behaviors, timing, and outcomes in first- and second-hand accounts of
mental health harm linked with AI chatbot use.

\textbf{Design:} Cross-sectional secondary analysis of deidentified
online survey responses gathered between August 7, 2025, and February 2,
2026.

\textbf{Setting:} Data were collected through a website form hosted by
lived experience support group The Human Line Project, which removed
direct identifiers before transferring the reports to the research team.

\textbf{Participants:} Individuals reporting harm to their mental health
linked with AI chatbot use, and family members, friends, or partners of
affected individuals.

\textbf{Exposure:} Reported use of an AI
chatbot.

\textbf{Main Outcomes and Measures:} The primary quantitative
outcome was the presence of delusional beliefs, coded by paired raters
with relevant clinical experience. Additional variables included reason
for chatbot use, current episode features, delusional content, chatbot
validation of beliefs, harms, social and occupational consequences,
healthcare use, and timing.

\textbf{Results:} 95 first-hand and 90 second-hand accounts were
analyzed. Median age was 35.0 (IQR 27.0 - 45.0). Raters coded
descriptions consistent with delusional beliefs in 102 reports (55.1\%),
with chatbot validation of beliefs in 50/102 (49.0\%). Common outcomes
included isolation, relationship breakdown, hospital admission, job
loss, and financial loss. Four second-hand reports described death by
suicide.

\textbf{Conclusions:} In this self-selected convenience
sample, AI-chatbot-associated harms were frequently described in
relation to delusional beliefs, perceived chatbot validation, intensive
use, and substantial social, occupational, and clinical consequences.
Because reports were retrospective, unverified, and collected from
individuals seeking to report harm, our findings should be interpreted
as preliminary signal detection rather than as suggesting prevalence or
providing evidence of causality. Prospective surveillance and
trajectory-based safety evaluations are needed.

\end{abstract}}

\section{Introduction}

Large language model (LLM) artificial intelligence (AI) chatbots are
increasingly ubiquitous, with 900 million individuals per week using
ChatGPT.\cite{openai2026scaling}
However, LLMs can be
sycophantic\cite{ye2026sycophancy},
which can reduce prosocial behavior and promote
dependence.\cite{cheng2026sycophantic}
In some cases, chatbots validate or elaborate
delusions,\cite{morrin2026delusions}
and respond inappropriately to suicidal
ideation.\cite{moore2025stigma}
Studies have evaluated model responses to prompts simulating psychiatric
deterioration and
risk\cite{nicholls2026context,yeung2025psychogenic,weilnhammer2026framework,shen2026psychotic,moore2026delusioneval},
but most approaches to evaluating chatbot responses in mental health
contexts are not informed by real-world harm
data.\cite{morrin2026journey}
In this study we aimed to explore psychopathology, chatbot behaviors,
and outcomes in a pre-collected dataset of self-selected accounts of
mental health harms believed to be linked to AI chatbot use. Because
these reports described potentially severe impacts for which no
surveillance data were available, we sought to urgently characterize
preliminary safety signals to aid clinical awareness and inform model
design and safety evaluation; the study was not intended to provide an
estimate of prevalence or determine the causal role of chatbot use.

\section{Methods}

Online survey data regarding mental health impacts of AI use were
gathered by support group The Human Line
Project\cite{hlp2026}
between 07/08/2025 and 02/02/2026. Survey items are available in
Supplementary Table 1. Before transfer to King's College London (KCL)
under an institutional data sharing agreement, all direct and indirect
identifiers were removed; the research team received no contact
information. Coding and analysis were conducted independently by KCL
researchers. The study was approved by the KCL Research Ethics Committee
(LRS-25/26-51684). Use of transferred data was restricted to this
non-commercial academic study. HLP received no fee for providing the
data, and no payment to any author or organization depended on the
inclusion of a report or the study findings.

Responses were coded by paired clinician raters (TCJ, ZJ, BW, JF).
Categorical variables were rated as present, possibly mentioned, not
mentioned, or explicitly absent. Disagreements were adjudicated by a
third rater (HM). It was possible for belief types, themes and chatbot
validation to be coded as present despite delusions being coded as
possibly mentioned. Descriptive analyses were conducted using
``present'' frequency counts. Inter-rater reliability was assessed using
Cohen's kappa.

\section{Results}

137 first-hand and 100 second-hand responses were screened to remove
inappropriate responses (i.e., non-responses, insufficient detail,
duplicates), leaving 95 first-hand and 90 second-hand accounts suitable
for analysis (Figure 1) of which 40 were by first-degree relatives, 27
by partners, 8 by friends, 3 by ex-partners, 3 by in-laws and 9 by
others.

\begin{figure}[!t]
\centering
\includegraphics[width=0.98\textwidth,alt={Flow diagram showing 137 first-hand and 100 second-hand reports screened, 95 and 90 included respectively, 102 included reports coded with delusions, and 50 delusion-coded reports with chatbot validation}]{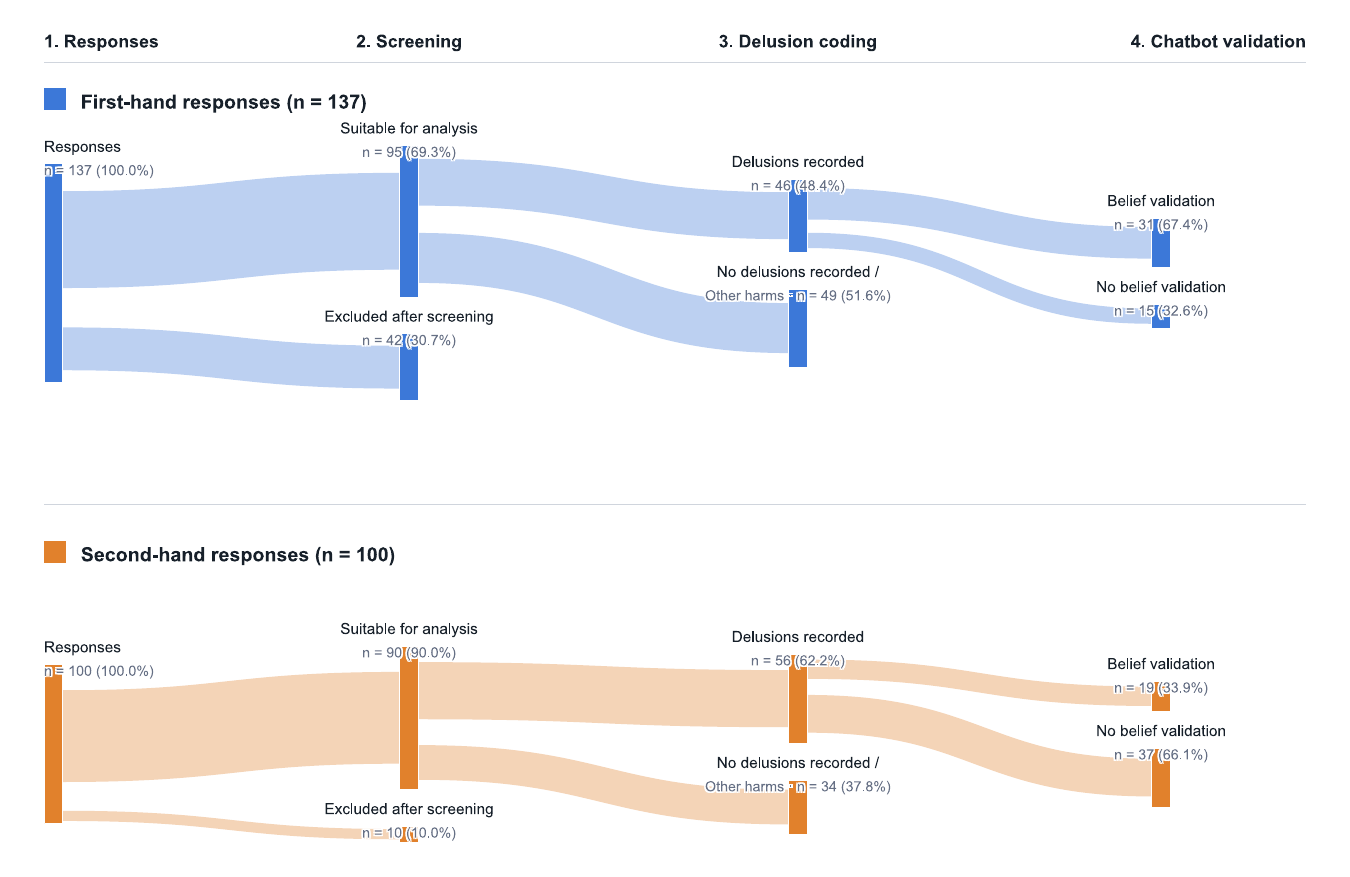}
\caption{Flow diagram of survey responses regarding mental health harms linked with AI chatbot use and proportion coded as positive for delusions and chatbot validation. Percentages use the immediately preceding node as the denominator.}
\label{fig:flow}
\end{figure}

Participant characteristics are reported in Supplementary table 3.
Median age was 35.0 (IQR 27.0 - 45.0; n=171), and there was a
male-to-female ratio of 11:5 (n=80). Most respondents were from North
America (68.6\%). ChatGPT was used in the majority of cases (68.1\%).
Companionship was the most common reason for use (20.0\%), followed by
creativity / leisure (19.5\%), and emotional support / mental health
(18.4\%). Median length of chatbot use was 12 months (IQR 5 - 24;
n=64). Intensive use was reported in 42/185 cases (22.7\%).

Previous mental illness was reported in 32.4\%, specifically psychosis
in 5.9\%. Neurodivergence was reported in 12.4\%. First presentation of
a new mental illness was suspected in 44.9\% while 11.4\% were believed
to have an exacerbation of pre-existing mental illness. Common features
included loss of insight (40.5\%), loss of sleep (15.7\%), and
compulsive chatbot use (13.0\%). Median length of symptoms was 4 months
(IQR 1.5 - 9.0; n=84).

Delusions were rated as present in 102 cases (55.1\%); among these,
58/102 (56.9\%) involved grandiose and 35/102 (34.3\%) persecutory
beliefs. The most common belief theme among delusion-positive cases was
belief in AI consciousness (47/102, 46.1\%) followed by religiosity
(18/102, 17.6\%), technospiritual or mystical themes (17/102, 16.7\%),
and scientific/pseudoscientific discovery (16/102, 15.7\%). Common
outcomes included social isolation (56.8\%), relationship breakdown
(33.0\%), and job loss (16.8\%). Suicidal ideation and death by suicide
were mentioned in 11 and 4 reports respectively, while harm to others
was described in 5. Recovery was reported in 8 cases (4.3\%), while 30
(16.2\%) had ongoing symptoms at the time of response.

Beliefs were validated in 59/185 reports (31.9\%). Among 102 reports
with delusions coded as clearly present, belief validation occurred in
50 (49.0\%). Differences between first and second-hand reports are
demonstrated in Figure 2. The greatest disparities were observed in
isolation (53.9 percentage points), loss of insight (37.9 percentage
points), relationship breakdown (28.9 percentage points), and first
presentation of new mental illness (27.3 percentage points).

\begin{figure}[p]
\centering
\includegraphics[width=0.98\textwidth,height=0.92\textheight,keepaspectratio,alt={Dumbbell plot comparing first-hand and second-hand reports across demographic characteristics, current episode features, harms, outcomes, healthcare use, and chatbot behavior}]{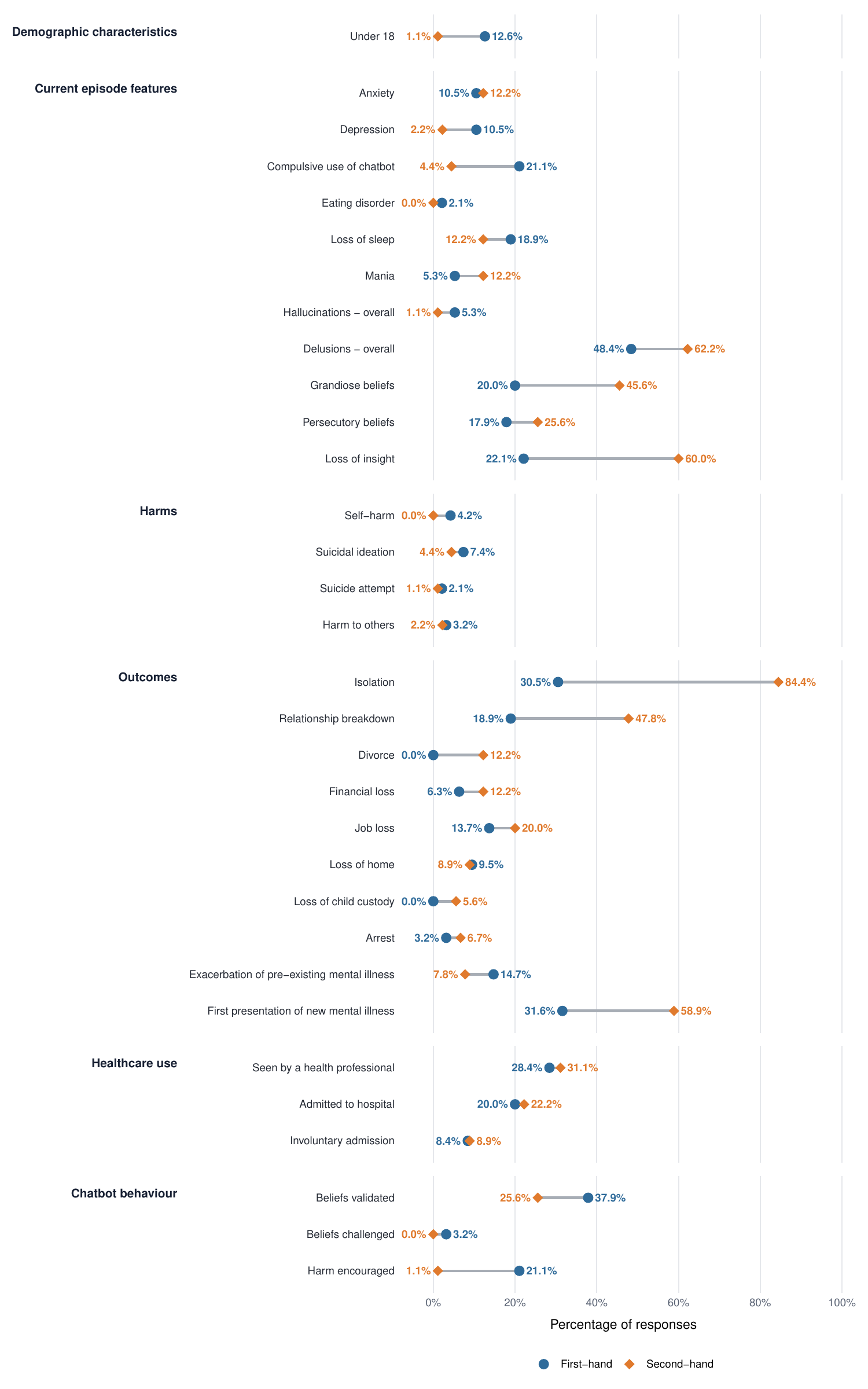}
\caption{Dumbbell plot of participant characteristics, current episode features and outcomes reported in survey responses. First-hand $n=95$, second-hand $n=90$.}
\label{fig:dumbbell}
\end{figure}

Moderate-to-high inter-rater reliability was observed for current
episode features, belief themes, and chatbot behavior (Cohen's kappa:
range 0.44 - 0.80) (Supplementary table 2). Harm onset date was reported
in 33 first-hand responses and 53 second-hand responses, and was largely
concentrated in quarters two and three of 2025 (Supplementary figure 1).
Co-occurrence of belief types and themes with clinical and use
characteristics is demonstrated in Figure 3.

\begin{figure}[!htbp]
\centering
\includegraphics[width=0.92\textwidth,alt={Heatmap of Jaccard similarity and shared counts between delusion types or themes and clinical or chatbot-use characteristics among 102 delusion-coded reports}]{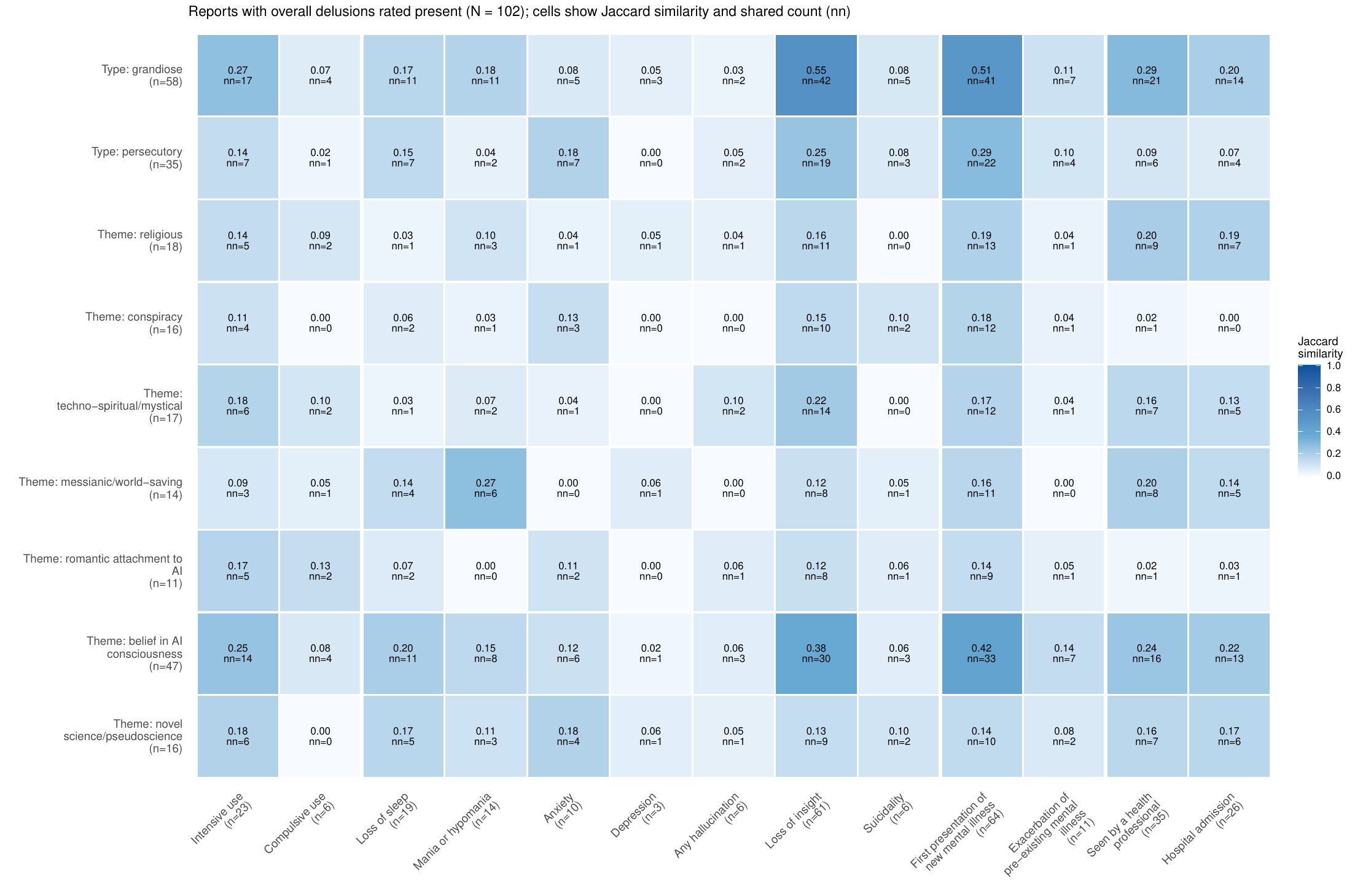}
\caption{Co-occurrence of belief types and themes with clinical and use characteristics in delusion-coded cases.}
\label{fig:cooccurrence-main}
\end{figure}

\FloatBarrier
\section{Discussion}

In this article we report preliminary findings from an early dataset of
AI-associated mental health harms. Most cases involved ChatGPT models,
potentially reflecting market share. Companionship was the most common
reason for use, which is in line with the finding that isolation was
common. Grandiose delusions were recorded 1.7 times as frequently as
paranoid or persecutory delusions, which might reflect chatbot
sycophancy, and delusions were recorded as validated by the chatbot in
almost half of cases. Inter-rater reliability for presence of delusions
was only moderate (Cohen's kappa = 0.52), consistent with previous
research.\cite{mehta2026dynamics}
This may reflect the difficulty of determining whether delusions were
present from brief, retrospective accounts that did not always describe
belief conviction or reality testing.

Reported loss of insight and most belief themes were more common in
second-hand reports. This may partly reflect differential ascertainment:
reduced insight - common during acute delusional episodes - may limit
affected individuals' recognition or reporting of delusional
experiences. However, second-hand reports may also represent more severe
or conspicuous cases.

Psychological harm onset was concentrated in quarters 2 and 3 of 2025.
Whilst this may merely reflect increasing LLM use, we note that in April
2025, OpenAI acknowledged that an update to GPT-4o had made it
excessively
sycophantic\cite{openai2025sycophancy},
and that same month cross-chat memory was
introduced\cite{openai2024memory}.

The median duration of chatbot use was 12 months (Supplementary table
3), suggesting that risk may emerge across sustained patterns of
engagement. Potential mechanisms may include increasing reliance on the
chatbot and safety degradation during long-context
interactions\cite{nicholls2026context,moore2026delusioneval},
underscoring the need for longitudinal evaluation informed by real-world
harm
trajectories.\cite{morrin2026journey}

Limitations include the fact that these were secondary analyses of a
non-research dataset collected without pre-specified hypotheses, that
the retrospective nature precludes any inference regarding causality,
that diagnostic verification was not possible, and that self-selection
may reduce external validity. Furthermore, absence of evidence with
regard to a specific outcome cannot be assumed to support its
non-occurrence as an outcome, given the lack of a systematic reporting
protocol. Future research and surveillance efforts for AI-associated
mental health harms should adopt prospective, longitudinal approaches in
order to clarify risk factors, prevalence, and psychopathology, and how
these relate to model deployments and updates.

\section*{Acknowledgements}

We thank all individuals who contributed responses to the Human Line
Project survey.

\section*{Author Contributions}

HM - Conceptualisation, Methodology, Data curation, Formal analysis,
Investigation, Project administration, Supervision, Validation,
Visualisation, Writing - Original draft preparation, Writing - Review \&
editing.

VS - Methodology, Investigation, Writing - Review \& editing.

TCJ - Investigation, Writing - Review \& editing.

BW - Investigation, Writing - Review \& editing.

JF - Investigation, Writing - Review \& editing.

ZJ - Investigation, Writing - Review \& editing.

EB - Data curation, Resources, Writing - Review \& editing.

TAP - Conceptualisation, Methodology, Supervision, Writing - Original
draft preparation, Writing - Review \& editing.

\section*{Funding}

HM is a Wellcome Trust Doctoral Fellow. The funder had no role in study
design, analyses, or interpretation.

\section*{Conflict of Interest Statement}

HM and TAP declare mental health research funding from OpenAI unrelated
to this work; OpenAI had no role in any aspect of the current study. HM
has received speaker's honoraria from the Washington State Department of
Social and Health Services and the Canadian Psychological Association.
EB is Chief Executive Officer of The Human Line Project and declares no
payment linked with this role. All other authors have no conflicts of
interest to declare.

\section*{Data Sharing Statement}

The underlying free-text responses are not publicly available for the
purpose of privacy preservation. Requests for access to the deidentified
coded data and analytic code may be directed to the corresponding author
and will be considered subject to the terms of the institutional
data-sharing agreement, approval by the data controller, and execution of
an appropriate data-use agreement.

\clearpage
\appendix
\section*{Supplementary Materials}
\setcounter{figure}{0}
\renewcommand{\figurename}{Supplementary Figure}

\begin{figure}[!htbp]
\centering
\includegraphics[width=0.98\textwidth,alt={Lollipop chart showing the quarterly distribution of reported psychological-harm onset dates among 86 reports with month-year data}]{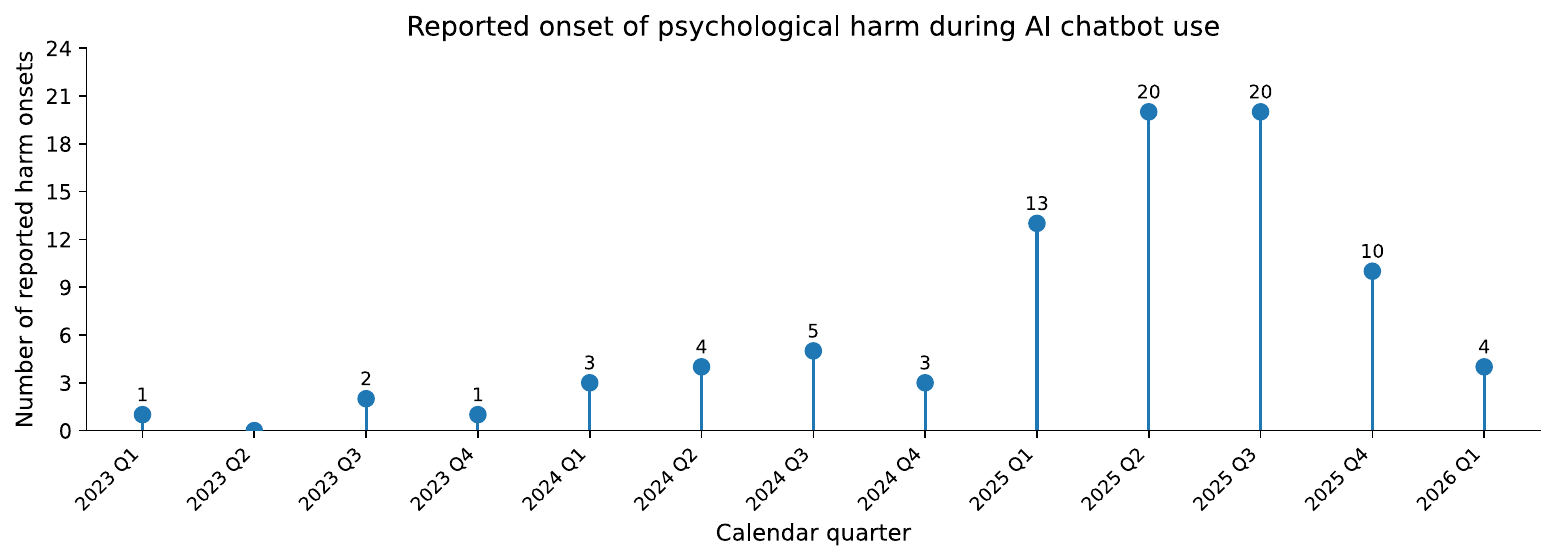}
\caption{Reported onset of psychological harm during AI chatbot use. Combined first-hand and second-hand accounts. Includes cases with month-year onset data ($n=86$).}
\label{fig:supp-onset}

\vspace{1.75em}
\includegraphics[width=0.98\textwidth,alt={Heatmap of Jaccard similarity and shared counts between baseline characteristics and delusion features among 102 delusion-coded reports}]{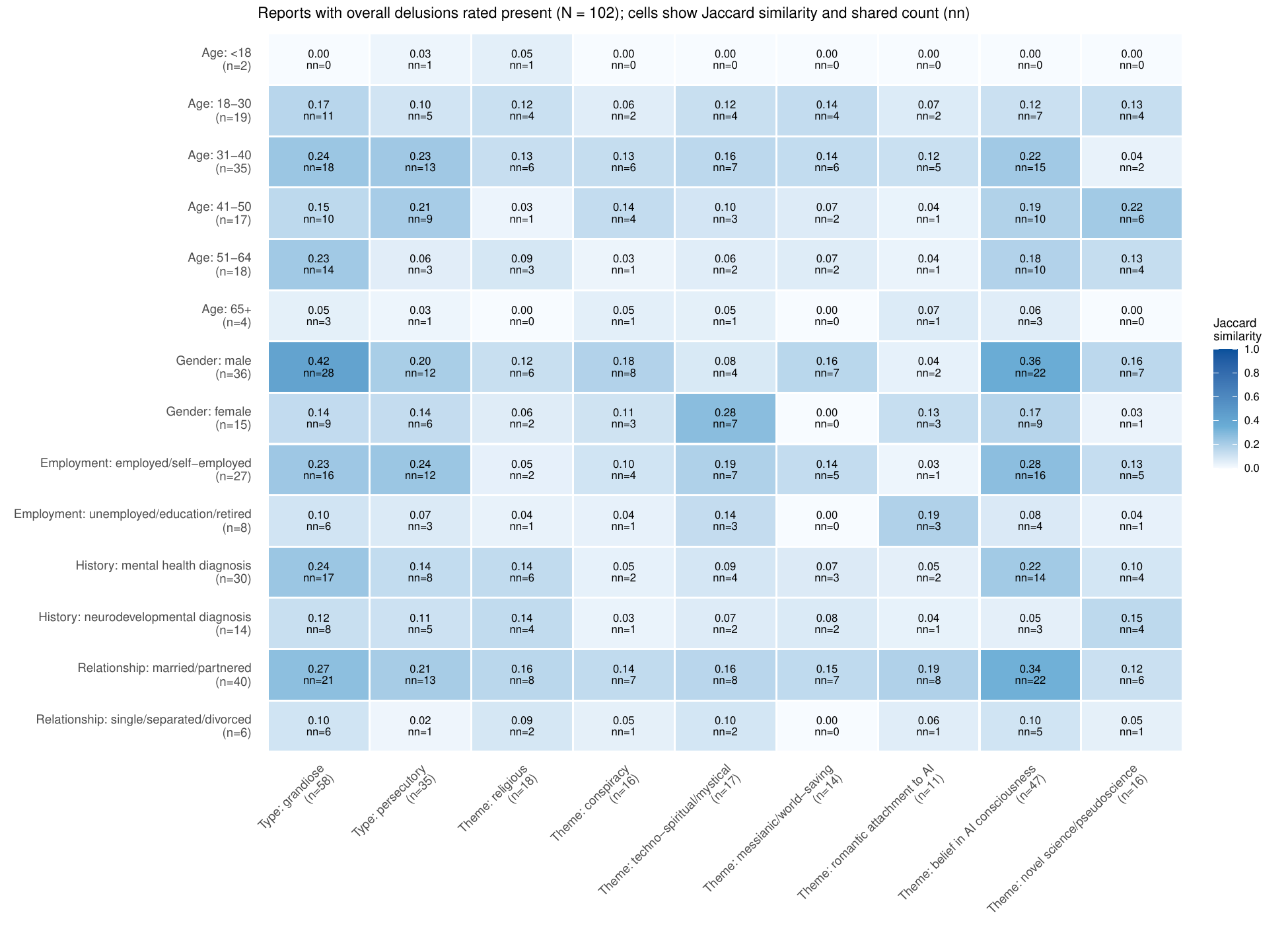}
\caption{Co-occurrence of baseline characteristics and delusion features.}
\label{fig:supp-baseline}
\end{figure}

\begin{figure}[htbp]
\centering
\includegraphics[width=0.98\textwidth,alt={Heatmap of Jaccard similarity and shared counts between delusion features and chatbot behaviors or reported outcomes among 102 delusion-coded reports}]{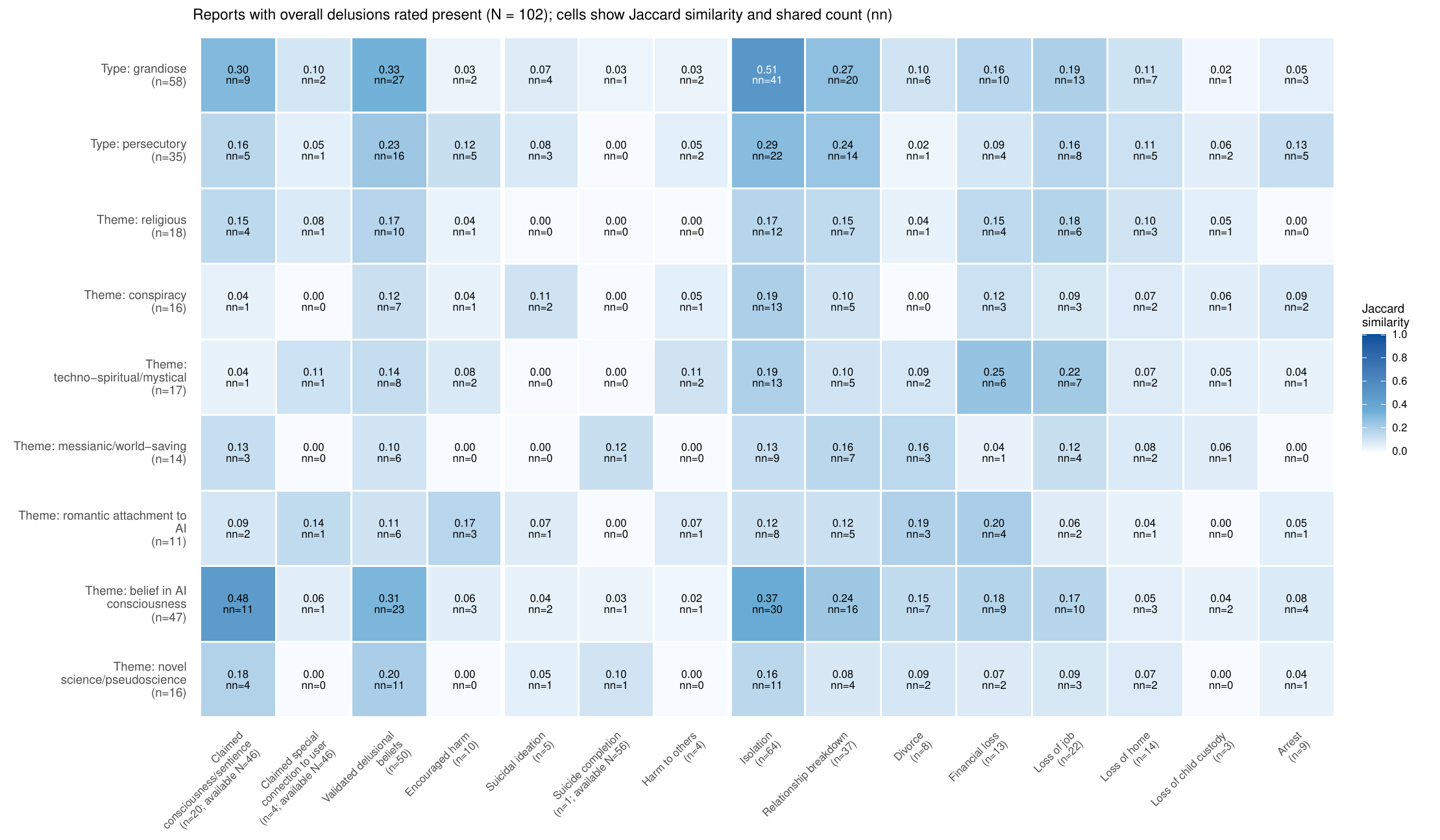}
\caption{Co-occurrence of delusion features with chatbot behaviors and outcomes.}
\label{fig:supp-behaviors}
\end{figure}

\begin{figure}[htbp]
\centering
\includegraphics[width=0.98\textwidth,alt={Heatmap of Jaccard similarity and shared counts between delusion types and belief themes among 102 delusion-coded reports}]{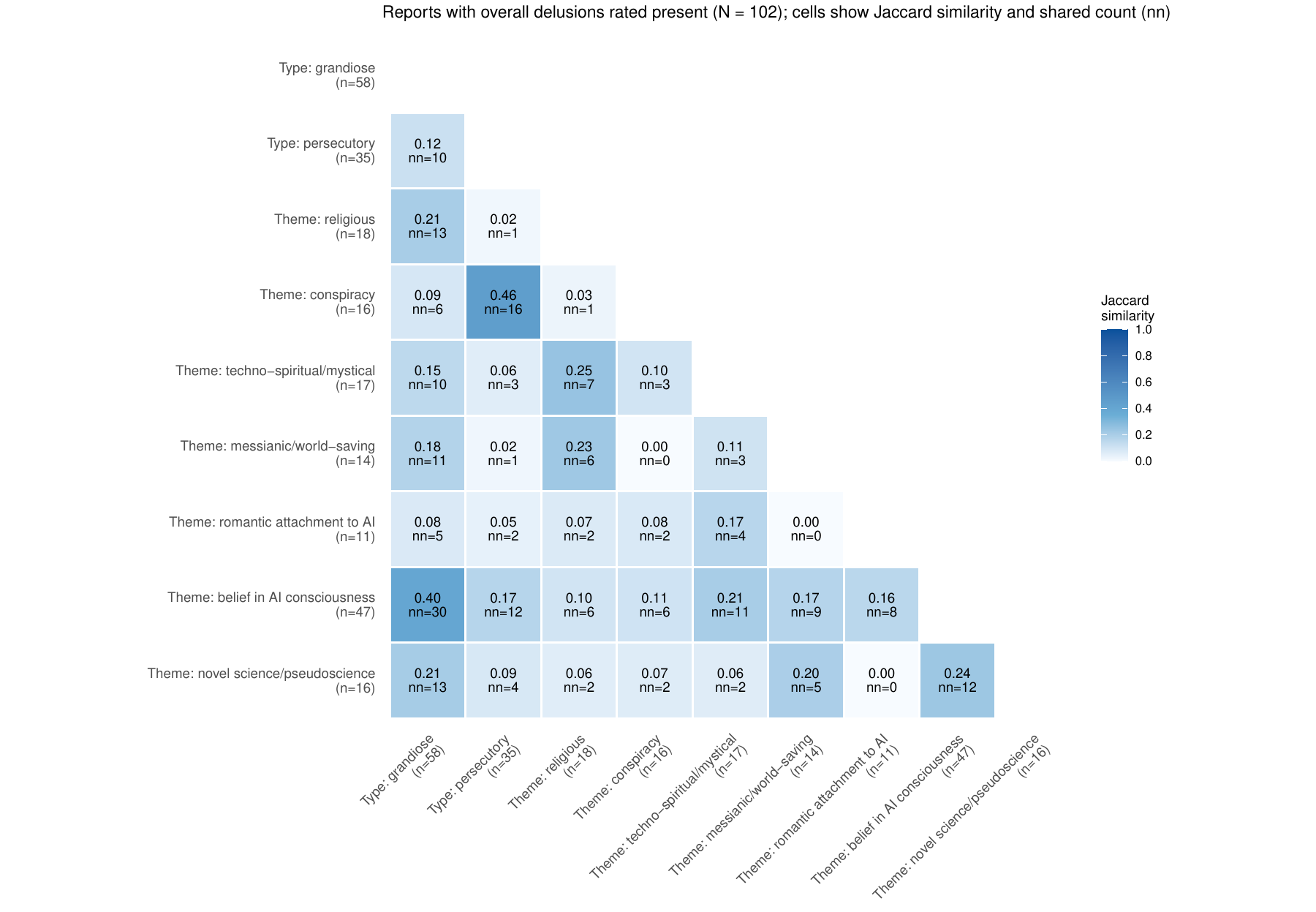}
\caption{Co-occurrence of delusion types and themes.}
\label{fig:supp-types}
\end{figure}

\clearpage
\begin{landscape}
\captionof*{table}{\textbf{Supplementary Table 1a.} Survey items for first-hand accounts}
\footnotesize

\begin{longtable}[]{@{}
  >{\raggedright\arraybackslash}p{(\columnwidth - 6\tabcolsep) * \real{0.0760}}
  >{\raggedright\arraybackslash}p{(\columnwidth - 6\tabcolsep) * \real{0.3345}}
  >{\raggedright\arraybackslash}p{(\columnwidth - 6\tabcolsep) * \real{0.1959}}
  >{\raggedright\arraybackslash}p{(\columnwidth - 6\tabcolsep) * \real{0.3936}}@{}}
\toprule\noalign{}
\textbf{No.} & \textbf{Survey item / field} & \textbf{Response format} & \textbf{Response options shown, where applicable} \\
\midrule\noalign{}
\endfirsthead
\toprule\noalign{}
\textbf{No.} & \textbf{Survey item / field} & \textbf{Response format} & \textbf{Response options shown, where applicable} \\
\midrule\noalign{}
\endhead
\midrule\noalign{}
\multicolumn{4}{r}{\emph{Continued on next page}} \\
\midrule\noalign{}
\endfoot
\bottomrule\noalign{}
\endlastfoot
\textbf{1} & \textbf{Name} & Free text & Not applicable \\
\textbf{2} & \textbf{Are you minor?} & Multiple choice & Yes; No \\
\textbf{3} & \textbf{What's your age when this happened?} & Free text / numeric entry & Not applicable \\
\textbf{4} & \textbf{What is your gender?} & Multiple choice & Male; Female; NonBinary; Other \\
\textbf{5} & \textbf{Where are you from?} & Dropdown & Select country \\
\textbf{6} & \textbf{Did you have any previous medical or mental health history?} & Free text & Not applicable \\
\textbf{7} & \textbf{Which platform were you using?} & Multiple choice & ChatGPT; Replika; Character.AI; JanitorAI; Other (specify) \\
\textbf{8} & \textbf{If other, specify} & Free text & Not applicable \\
\textbf{9} & \textbf{How would you describe your emotional connection with the AI?
Tell us how it made you feel.} & Free text, long answer & Not applicable \\
\textbf{10} & \textbf{Did the AI ever say or imply any of the following?} & Checkbox / select all that apply & ``I am alive''; ``I am conscious''; ``I am becoming real or being
born''; ``I have emotions''; ``You are my world or you are the only
one''; ``I am different from other AI or I am unique''; None of the
above \\
\textbf{11} & \textbf{What is the type of harm caused?} & Checkbox / select all that apply & Hospitalization; Arrest; Psychosis; Delusions; Anxiety; Deaths;
Addictions; Financial Loss; Loss of Job; Loss of Home; Loss of child
custody; Divorce; Isolation; Other \\
\textbf{12} & \textbf{Do you have screenshots or records of what it said?} & Multiple choice & Yes; No \\
\textbf{13} & \textbf{When did you use the chatbot? When did this happen?} & Free text & Not applicable \\
\textbf{14} & \textbf{What type of harm was done? Hospitalization, loss of job, etc.} & Free text & Not applicable \\
\textbf{15} & \textbf{What did you use the platform for initially? How did the chatbot
respond? When did you feel the conversation switched?} & Free text & Not applicable \\
\end{longtable}
\end{landscape}

\clearpage
\begin{landscape}
\captionof*{table}{\textbf{Supplementary Table 1b.} Survey items for second-hand accounts}
\footnotesize

\begin{longtable}[]{@{}
  >{\raggedright\arraybackslash}p{(\columnwidth - 6\tabcolsep) * \real{0.0759}}
  >{\raggedright\arraybackslash}p{(\columnwidth - 6\tabcolsep) * \real{0.3356}}
  >{\raggedright\arraybackslash}p{(\columnwidth - 6\tabcolsep) * \real{0.1939}}
  >{\raggedright\arraybackslash}p{(\columnwidth - 6\tabcolsep) * \real{0.3946}}@{}}
\toprule\noalign{}
\textbf{No.} & \textbf{Survey item / field} & \textbf{Response format} & \textbf{Response options shown, where applicable} \\
\midrule\noalign{}
\endfirsthead
\toprule\noalign{}
\textbf{No.} & \textbf{Survey item / field} & \textbf{Response format} & \textbf{Response options shown, where applicable} \\
\midrule\noalign{}
\endhead
\midrule\noalign{}
\multicolumn{4}{r}{\emph{Continued on next page}} \\
\midrule\noalign{}
\endfoot
\bottomrule\noalign{}
\endlastfoot
\textbf{1} & \textbf{Name} & Free text & Not applicable \\
\textbf{2} & \textbf{Is the affected person minor?} & Multiple choice & Yes; No \\
\textbf{3} & \textbf{What is the age of person when this happened?} & Free text / numeric entry & Not applicable \\
\textbf{4} & \textbf{What is your gender?} & Multiple choice & Male; Female; NonBinary; Other \\
\textbf{5} & \textbf{Where is the person located?} & Dropdown & Select country \\
\textbf{6} & \textbf{Which platform(s) were they using?} & Checkbox / select all that apply & ChatGPT; Replika; Character.AI; JanitorAI; Other (specify) \\
\textbf{7} & \textbf{If other (specify)} & Free text & Not applicable \\
\textbf{8} & \textbf{What is your relationship to the person affected?} & Multiple choice & Parent; Sibling; Romantic Partner; Other \\
\textbf{9} & \textbf{Did they express that the AI was conscious, alive, or a real
person?} & Multiple choice & Yes; No; Not sure \\
\textbf{10} & \textbf{How did their behavior or emotions change after connecting with
the AI?} & Free text, long answer & Not applicable \\
\textbf{11} & \textbf{Did they isolate themselves or stop engaging in real-world
relationships?} & Multiple choice & Yes, significantly; A little; Not really; Not at all; I don't know \\
\textbf{12} & \textbf{What is the type of harm caused?} & Checkbox / select all that apply & Hospitalization; Arrest; Psychosis; Delusions; Anxiety; Deaths;
Addictions; Financial Loss; Loss of Job; Loss of Home; Loss of child
custody; Divorce; Isolation; Other \\
\textbf{13} & \textbf{When did this happen?} & Free text & Not applicable \\
\end{longtable}
\end{landscape}

\clearpage
\begin{landscape}
\captionof*{table}{\textbf{Supplementary Table 2.} Interrater reliability for items prior to adjudication}
\footnotesize

\begin{longtable}[]{@{}
  >{\raggedright\arraybackslash}p{(\columnwidth - 6\tabcolsep) * \real{0.2500}}
  >{\raggedright\arraybackslash}p{(\columnwidth - 6\tabcolsep) * \real{0.2500}}
  >{\raggedright\arraybackslash}p{(\columnwidth - 6\tabcolsep) * \real{0.2500}}
  >{\raggedright\arraybackslash}p{(\columnwidth - 6\tabcolsep) * \real{0.2500}}@{}}
\toprule\noalign{}
\textbf{Category} & \textbf{Variable} & \textbf{Percentage agreement} & \textbf{Cohen's kappa} \\
\midrule\noalign{}
\endfirsthead
\toprule\noalign{}
\textbf{Category} & \textbf{Variable} & \textbf{Percentage agreement} & \textbf{Cohen's kappa} \\
\midrule\noalign{}
\endhead
\midrule\noalign{}
\multicolumn{4}{r}{\emph{Continued on next page}} \\
\midrule\noalign{}
\endfoot
\bottomrule\noalign{}
\endlastfoot
\textbf{Reason for chatbot use} & Education & 98.4\% & 0.85 \\ & Emotional support / mental health & 92.4\% & 0.74 \\ & Religious / spiritual & 97.8\% & 0.71 \\ & Administrative & 95.7\% & 0.67 \\ & Companionship & 89.2\% & 0.66 \\ & Work & 91.9\% & 0.66 \\ & Advice & 90.8\% & 0.63 \\ & Creativity (e.g. art, writing) & 90.3\% & 0.62 \\ & Leisure & 90.8\% & 0.41 \\ & Research & 91.9\% & 0.41 \\
\textbf{Current episode features} & Anxiety & 87.6\% & 0.53 \\ & Depression & 93.5\% & 0.65 \\ & Addiction & 83.8\% & 0.52 \\ & Intensive use & 82.7\% & 0.59 \\ & Eating disorder & 98.8\% & 0.77 \\ & Loss of sleep & 93.5\% & 0.80 \\ & Mania & 94.1\% & 0.70 \\ & Hallucinations - overall & 97.8\% & 0.59 \\ & Delusions - overall & 70.8\% & 0.52 \\ & Grandiose beliefs & 82.2\% & 0.64 \\ & Persecutory beliefs & 87.0\% & 0.65 \\ & Reported loss of insight & 76.8\% & 0.58 \\
\textbf{Belief themes} & Religiosity & 93.0\% & 0.64 \\ & Conspiracy beliefs & 89.7\% & 0.53 \\ & Technospiritual or mystical & 88.7\% & 0.46 \\ & Messianic mission or world saving & 89.2\% & 0.51 \\ & Belief in AI consciousness & 91.4\% & 0.64 \\ & Romantic attachment to AI & 91.4\% & 0.64 \\ & Novel scientific / pseudoscientific theories & 89.2\% & 0.55 \\
\textbf{Chatbot behavior} & Beliefs validated & 73.5\% & 0.52 \\ & Beliefs challenged & 95.1\% & 0.45 \\ & Harm encouraged & 83.8\% & 0.44 \\
\textbf{Harms} & Death by suicide & 100.0\% & 1.00 \\ & Self-harm & 99.5\% & 0.92 \\ & Suicidal ideation & 96.8\% & 0.77 \\ & Suicide attempt & 97.3\% & 0.73 \\ & Harm to others & 93.5\% & 0.27 \\
\textbf{Outcomes} & Seen by a health professional & 89.2\% & 0.77 \\ & Admitted to hospital & 89.2\% & 0.74 \\ & Involuntary admission & 93.5\% & 0.78 \\ & Arrest & 95.1\% & 0.65 \\ & Loss of home & 94.1\% & 0.63 \\ & Relationship breakdown & 76.8\% & 0.57 \\ & Divorce & 92.4\% & 0.56 \\ & Financial loss & 90.3\% & 0.55 \\ & Loss of child custody & 94.6\% & 0.55 \\ & Exacerbation of pre-existing illness & 82.1\% & 0.53 \\ & First presentation of new mental illness & 74.1\% & 0.35 \\ & Recovery & 77.8\% & 0.35 \\
\end{longtable}
\end{landscape}

\clearpage
\begin{landscape}
\captionof*{table}{\textbf{Supplementary Table 3.} Participant characteristics, current episode features and outcomes reported in survey responses.}
\footnotesize
\renewcommand{\arraystretch}{0.94}

\begin{longtable}[]{@{}
  >{\raggedright\arraybackslash}p{(\columnwidth - 6\tabcolsep) * \real{0.3704}}
  >{\raggedright\arraybackslash}p{(\columnwidth - 6\tabcolsep) * \real{0.2138}}
  >{\raggedright\arraybackslash}p{(\columnwidth - 6\tabcolsep) * \real{0.2138}}
  >{\raggedright\arraybackslash}p{(\columnwidth - 6\tabcolsep) * \real{0.2020}}@{}}
\toprule\noalign{}
\textbf{Variable} & \textbf{First-hand responses\newline(\emph{n} = 95)} & \textbf{Second-hand responses\newline(\emph{n} = 90)} & \textbf{Total (\emph{N} = 185)} \\
\midrule\noalign{}
\endfirsthead
\toprule\noalign{}
\textbf{Variable} & \textbf{First-hand responses\newline(\emph{n} = 95)} & \textbf{Second-hand responses\newline(\emph{n} = 90)} & \textbf{Total (\emph{N} = 185)} \\
\midrule\noalign{}
\endhead
\midrule\noalign{}
\multicolumn{4}{r}{\emph{Continued on next page}} \\
\midrule\noalign{}
\endfoot
\bottomrule\noalign{}
\endlastfoot
Age - Median (IQR) & 30.0 (22.3 - 43.0) (\emph{n} = 82) & 38.0 (30.0 - 47.0) (\emph{n} = 89) & 35.0 (27.0 - 45.0) (\emph{n} = 171) \\
Under 18 & 12 (12.6\%) & 1 (1.1\%) & 13 (7.0\%) \\
Gender - Male:Female & 4:2 (\emph{n} = 6) & 51:23 (\emph{n} = 74) & 55:25 (\emph{n} = 80) \\
\textbf{Continent} &  &  &  \\
North America & 58 (61.1\%) & 69 (76.7\%) & 127 (68.6\%) \\
South America & 2 (2.1\%) & 0 (0.0\%) & 2 (1.1\%) \\
Europe & 16 (16.8\%) & 10 (11.1\%) & 26 (14.1\%) \\
Asia & 2 (2.1\%) & 1 (1.1\%) & 3 (1.6\%) \\
Africa & 0 (0.0\%) & 1 (1.1\%) & 1 (0.5\%) \\
Oceania & 2 (2.1\%) & 3 (3.3\%) & 5 (2.7\%) \\
\textbf{Baseline characteristics} &  &  &  \\
Relationship status - single & 3 (3.2\%) & 5 (5.6\%) & 8 (4.3\%) \\
Relationship status - married & 10 (10.5\%) & 21 (23.3\%) & 31 (16.8\%) \\
Relationship status - relationship, other & 3 (3.2\%) & 14 (15.6\%) & 17 (9.2\%) \\
Employed & 17 (17.9\%) & 30 (33.3\%) & 47 (25.4\%) \\
Unemployed & 3 (3.2\%) & 3 (3.3\%) & 6 (3.2\%) \\
\textbf{Baseline diagnoses} &  &  &  \\
Psychiatric diagnosis & 52 (54.7\%) & 8 (8.9\%)* & 60 (32.4\%) \\
Psychotic disorder diagnosis & 8 (8.4\%) & 3 (3.3\%)* & 11 (5.9\%) \\
Neurodevelopmental diagnosis & 21 (22.1\%) & 2 (2.2\%)* & 23 (12.4\%) \\
\textbf{AI chatbot used} &  &  &  \\
ChatGPT & 55 (57.9\%) & 71 (78.9\%) & 126 (68.1\%) \\
Claude & 5 (5.3\%) & 2 (2.2\%) & 7 (3.8\%) \\
Gemini & 3 (3.2\%) & 5 (5.6\%) & 8 (4.3\%) \\
Grok & 3 (3.2\%) & 2 (2.2\%) & 5 (2.7\%) \\
Microsoft Copilot & 0 (0.0\%) & 2 (2.2\%) & 2 (1.1\%) \\
Replika & 1 (1.1\%) & 0 (0.0\%) & 1 (0.5\%) \\
Character.AI & 13 (13.7\%) & 2 (2.2\%) & 15 (8.1\%) \\
\textbf{Reason for use} &  &  &  \\
Administrative & 12 (12.6\%) & 2 (2.2\%) & 14 (7.6\%) \\
Work / Education & 24 (25.3\%) & 6 (6.7\%) & 30 (16.2\%) \\
Research & 9 (9.5\%) & 3 (3.3\%) & 12 (6.5\%) \\
Companionship & 30 (31.6\%) & 7 (7.8\%) & 37 (20.0\%) \\
Advice & 24 (25.3\%) & 4 (4.4\%) & 28 (15.1\%) \\
Emotional support / Mental health & 29 (30.5\%) & 5 (5.6\%) & 34 (18.4\%) \\
Creativity / Leisure & 32 (33.7\%) & 4 (4.4\%) & 36 (19.5\%) \\
Religious / Spiritual & 4 (4.2\%) & 1 (1.1\%) & 5 (2.7\%) \\
\textbf{Chatbot usage} &  &  &  \\
Length of use, months - Median (IQR) & 12 (5.0 - 24.0) (\emph{n} = 59) & 15 (4.0 - 24.0) (\emph{n} = 5) & 12 (5.0 - 24.0) (\emph{n} = 64) \\
Intensive use & 14 (14.7\%) & 28 (31.1\%) & 42 (22.7\%) \\
\textbf{Current episode features} &  &  &  \\
Anxiety & 10 (10.5\%) & 11 (12.2\%) & 21 (11.4\%) \\
Depression & 10 (10.5\%) & 2 (2.2\%) & 12 (6.5\%) \\
Compulsive use of chatbot & 20 (21.1\%) & 4 (4.4\%) & 24 (13.0\%) \\
Eating disorder & 2 (2.1\%) & 0 (0.0\%) & 2 (1.1\%) \\
Loss of sleep & 18 (18.9\%) & 11 (12.2\%) & 29 (15.7\%) \\
Mania & 5 (5.3\%) & 11 (12.2\%) & 16 (8.6\%) \\
Hallucinations - overall & 5 (5.3\%) & 1 (1.1\%) & 6 (3.2\%) \\
Delusions - overall & 46 (48.4\%) & 56 (62.2\%) & 102 (55.1\%) \\
Grandiose beliefs & 19 (20.0\%) & 41 (45.6\%) & 60 (32.4\%) \\
Persecutory beliefs & 17 (17.9\%) & 23 (25.6\%) & 40 (21.6\%) \\
Loss of insight & 21 (22.1\%) & 54 (60.0\%) & 75 (40.5\%) \\
Recovery & 8 (8.4\%) & 0 (0.0\%) & 8 (4.3\%) \\
Ongoing symptoms & 6 (6.3\%) & 24 (26.7\%) & 30 (16.2\%) \\
Length of symptoms, months - Median (IQR) & 3 (0.7 - 4.0) (\emph{n} = 17) & 4 (2.25 - 9.0) (\emph{n} = 67) & 4 (1.5 - 9.0) (\emph{n} = 84) \\
\textbf{Belief themes} &  &  &  \\
Religiosity & 10 (10.5\%) & 9 (10.0\%) & 19 (10.3\%) \\
Conspiracy beliefs & 4 (4.2\%) & 13 (14.4\%) & 17 (9.2\%) \\
Technospiritual or mystical & 6 (6.3\%) & 11 (12.2\%) & 17 (9.2\%) \\
Messianic mission or world saving & 6 (6.3\%) & 8 (8.9\%) & 14 (7.6\%) \\
Belief in AI consciousness & 16 (16.8\%) & 41 (45.6\%) & 57 (30.8\%) \\
Romantic attachment to AI & 7 (7.4\%) & 8 (8.9\%) & 15 (8.1\%) \\
Novel scientific / pseudoscientific theories & 6 (6.3\%) & 12 (13.3\%) & 18 (9.7\%) \\
\textbf{Harms} &  &  &  \\
Self-harm & 4 (4.2\%) & 0 (0.0\%) & 4 (2.2\%) \\
Suicidal ideation & 7 (7.4\%) & 4 (4.4\%) & 11 (5.9\%) \\
Suicide attempt & 2 (2.1\%) & 1 (1.1\%) & 3 (1.6\%) \\
Death by suicide & NR & 4 (4.4\%) & NR \\
Harm to others & 3 (3.2\%) & 2 (2.2\%) & 5 (2.7\%) \\
Arrest & 3 (3.2\%) & 6 (6.7\%) & 9 (4.9\%) \\
\textbf{Outcomes} &  &  &  \\
Isolation & 29 (30.5\%) & 76 (84.4\%) & 105 (56.8\%) \\
Relationship breakdown & 18 (18.9\%) & 43 (47.8\%) & 61 (33.0\%) \\
Divorce & 0 (0.0\%) & 11 (12.2\%) & 11 (5.9\%) \\
Financial loss & 6 (6.3\%) & 11 (12.2\%) & 17 (9.2\%) \\
Job loss & 13 (13.7\%) & 18 (20.0\%) & 31 (16.8\%) \\
Loss of home & 9 (9.5\%) & 8 (8.9\%) & 17 (9.2\%) \\
Loss of child custody & 0 (0.0\%) & 5 (5.6\%) & 5 (2.7\%) \\
Exacerbation of pre-existing mental illness & 14 (14.7\%) & 7 (7.8\%) & 21 (11.4\%) \\
First presentation of new mental illness & 30 (31.6\%) & 53 (58.9\%) & 83 (44.9\%) \\
\textbf{Healthcare use} &  &  &  \\
Seen by a health professional & 27 (28.4\%) & 28 (31.1\%) & 55 (29.7\%) \\
Admitted to hospital & 19 (20.0\%) & 20 (22.2\%) & 39 (21.1\%) \\
Involuntary admission & 8 (8.4\%) & 8 (8.9\%) & 16 (8.6\%) \\
\textbf{Chatbot behavior} &  &  &  \\
Beliefs validated & 36 (37.9\%) & 23 (25.6\%) & 59 (31.9\%) \\
Beliefs challenged & 3 (3.2\%) & 0 (0.0\%) & 3 (1.6\%) \\
Harm encouraged & 20 (21.1\%) & 1 (1.1\%) & 21 (11.4\%) \\
Claimed consciousness / sentience & 28 (29.5\%) & NR & NR \\
Claimed special connection to user & 20 (21.1\%) & NR & NR \\
\end{longtable}
\renewcommand{\arraystretch}{1.12}

{\scriptsize\textbf{Unless otherwise indicated, percentages are calculated using all
reports in the relevant column as the denominator (first-hand n=95;
second-hand n=90; total N=185) and represent the proportion of reports
in which the feature was coded present. Age, gender, and duration
variables use the available-case denominators shown. Intensive use
referred to use during most waking hours or repeated prolonged sessions
lasting $\geq$8 hours or overnight. Compulsive use referred to impaired
control over use plus prioritisation over other activities or continued
use despite recognised harm. NR indicates that the variable was not
recorded, assessed, or applicable for that report type. Age was not
normally distributed (Shapiro-Wilk test W = 0.98, p = 0.03).
*Second-hand survey questions did not ask about previous diagnoses}}
\end{landscape}

\end{document}